\documentclass[%
 reprint,
 amsmath,amssymb,
 aps,
]{revtex4-2}
\usepackage{physics}
\usepackage{graphicx}
\usepackage{dcolumn}
\usepackage{bm}

\usepackage{amsmath}
\usepackage{graphicx}
\usepackage[colorinlistoftodos]{todonotes}
\usepackage[colorlinks=true, allcolors=blue]{hyperref}
\usepackage{xfrac}
\usepackage{units}
\usepackage{amssymb}
\usepackage{dsfont}
\usepackage{epstopdf}
\usepackage{tikz}
\usetikzlibrary{calc}
\usepackage{nicefrac}
\usepackage{mathtools}
\usepackage{bm}
\usepackage{epstopdf}
\usepackage{float}
\usepackage{tabularx}
\usepackage[labelformat=simple]{subcaption}

\DeclarePairedDelimiter{\floor}{\lfloor}{\rfloor}
\newcommand{\RomanNumeralCaps}[1]
    {\MakeUppercase{\romannumeral #1}}

\begin{document}

\preprint{APS/123-QED}

\title{Degeneracy and coherent states of the two-dimensional Morse potential}

\author{James Moran}
 \email{james.moran@umontreal.ca}
\affiliation{D\'epartement de physique, Universit\'e de Montr\'eal,\\ Montr\'eal, Qu\'ebec, H3C 3J7, Canada}%

\affiliation{Centre de recherches math\'ematiques, Universit\'e de Montr\'eal,\\ Montr\'eal, Qu\'ebec, H3C 3J7, Canada}%

\begin{abstract}
In this paper we construct coherent states for the two-dimensional Morse potential. We find the dependence of the spectrum on the physical parameters and use this to understand the emergence of accidental degeneracies. It is observed that, under certain conditions pertaining to the irrationality of the parameters, accidental degeneracies do not appear and as such energy levels are at most two-fold degenerate. After defining a non-degenerate spectrum and set of states for the 2D Morse potential, we construct generalised coherent states and discuss the spatial distribution of their probability densities and their uncertainty relations.
\end{abstract}

\maketitle


\section{Introduction}
The Morse potential was originally introduced as a means to model interactions in diatomic molecules \cite{PhysRev.34.57}, it is an exactly solvable model with eigenfunctions expressible in terms of Laguerre polynomials. In physical contexts the two-dimensional Morse product eigenfunctions have been used as a basis for perturbative solutions to a triatomic molecular hamiltonian  \cite{doi:10.1063/1.1670777, doi:10.1063/1.1670678, doi:10.1063/1.2140500, https://doi.org/10.1002/qua.21189, Apanavicius:2021yin}. Throughout physics the Morse potential is used in a variety of applications including the study of graphene \cite{ZALI2021413045, C3CP55340J}, spectroscopy \cite{doi:10.1063/5.0031216, 10.1088/978-0-7503-1989-8ch12}, Bose-Einstein condensation \cite{PhysRevA.91.023604}, theories of interacting electrons \cite{PhysRevE.95.052217}, nuclear physics \cite{PhysRevC.84.024309}, supersymmetric quantum mechanics \cite{doi:10.1142/S0217751X1250073X}, and molecular dynamics \cite{doi:10.1063/1.5041911}. Coherent states for the 1D Morse potential have been studied \cite{Angelova_2008}, and while there is literature on defining coherent states for systems with degenerate spectra \cite{PhysRevA.64.042104}, and coherent states for the 2D square well have been analysed \cite{Fiset_2015}, so far coherent states for the 2D Morse potential have not been explicitly defined.

The 2D Morse potential, and more specifically, its supersymmetric generalisations have been studied in detail \cite{Ioffe_2004, IOFFE20062552, Ioffe_2005, PhysRevA.76.052114, ISI:000288078700002}. Because the supersymmetric partners of the 2D Morse share the same spectrum, they also give rise to accidentally degenerate states, the existence of which may be explained in terms of an operator constructed from supercharges \cite{ISI:000288078700002}. For our purposes we do not need to invoke the framework of supersymmetry, instead we can study the degeneracy in terms of the rationality of the physical constants appearing in the definition of the potential. 

Degeneracy arises in practically all multidimensional quantum systems and in the Morse potential the degeneracy is found to be quadratic in the principle quantum number. Categorising degeneracies in 2D systems with quadratic spectra has solutions found in number theory. This is distinct from the case where, for example, the spectrum is linear in the quantum number, we can associate a known symmetry group to the spectrum such as $U(n)$ for the $n$-dimensional isotropic oscillator \cite{PhysRev.103.1119}.

There are several quantum systems with quadratic spectra, including the Morse potential, P\"oschl-Teller \cite{Poschl1933} and the particle in an infinite square box \cite{Sakurai2017}. The prototypical 2D quadratic spectrum is that of the particle in a square box. Finding the degenerate energies is equivalent to finding numbers which are the sum of the squares of two integers \cite{Shaw_1974}. Because the square box only admits bound states, in principle we need to find solutions to the sum of the squares of two integers to infinity.

The 2D Morse spectrum is different in a few key ways. Firstly, if we are just interested in the bound states of the system, we need only to find energies up to a certain finite value. Secondly, the behaviour of the degeneracies changes when the defining parameter in the Morse potential changes. As we will show, accidentally degenerate solutions do not exist when this parameter is irrational.

The understanding of the degeneracy of multidimensional quantum systems is important in the construction of generalised coherent states \cite{PhysRevA.64.042104}. Indeed, the usual definitions require a spectrum to be ordered as
\begin{equation}
E_0<E_1<E_2<\ldots<E_M,
\end{equation}
in order to fulfil the resolution of the identity \cite{Klauder_1996}. For most systems it is important to work with complete sets of states, in the Morse potential however, we will only be studying the finite dimensional bound state spectrum and as a result do not require a resolution of the identity on the entire Hilbert space, though one in principle may be constructed on the finite dimensional bound state sector. This being said, the ordering of the spectrum is still required so that we can extend the formalism of generalised coherent states to the 2D Morse potential without modifying the existing definitions. 

The paper is organised as follows. Firstly, in section \ref{introduction} we define the energy eigenstates and eigenvalues of the 2D Morse potential as well its principle parameter, $p$. Following this in section \ref{sec3} we discuss the nature of the degeneracies that may arise depending on the rationality of the principle parameter. Focussing on the case of irrational $p$, in section \ref{sec4} we introduce two parameters, $\gamma, \delta$ which control the mixing of the degenerate contributions and define the cumulative wavefunctions. Lastly, in sections \ref{sec5} and \ref{sec6} we assess the behaviour of the generalised coherent states and their probability distributions as well as computing their uncertainty relations, we conclude by discussing future work that could be made on the subject.

\section{The 2D Morse potential and the parameter $p$}\label{introduction}
The 2D isotropic Morse Hamiltonian is defined by
\begin{equation}\label{ham}
\begin{split}
\hat{H}=&\frac{1}{2m}\left(\hat{P}_x^2 +\hat{P}_y^2\right)\\
& +V_0\left(e^{-2\beta \hat{Q}_x}+e^{-2\beta \hat{Q}_y}-2\left(e^{-\beta \hat{Q}_x}+e^{-\beta \hat{Q}_y}\right) \right),
\end{split}
\end{equation}
where $\hat{P}_x, \hat{P}_y$ are the momentum operators and $\hat{Q}_x, \hat{Q}_y$ are the corresponding position operators. The Hamiltonian \eqref{ham} is isotropic in the sense that the parameters in the $x$ mode are equal to those in the $y$ mode. The entire Hilbert space decomposes into the sum of a finite dimensional bound state part with discrete spectrum and an infinite dimensional unbound state part with continuous spectrum. In the present work we are concerned with only the bound states of the Morse oscillator.

The bound quantum states are found by solving the stationary Schr\"odinger equation
\begin{equation}\label{eval}
\hat{H}\ket{n,m}=E_{n,m}\ket{n,m},
\end{equation}
to obtain energy eigenvalues
\begin{equation}\label{energy}
E_{n,m}=-\frac{\hbar^2 \beta^2}{2m}\left(\left( p-n\right)^2 +\left( p-m\right)^2\right)
\end{equation}
where the parameters
\begin{equation}\label{mainparam}
\nu=\sqrt{\frac{8mV_0}{\beta^2 \hbar^2}}, \quad p=\frac{\nu-1}{2},
\end{equation}
have been defined. The parameter $p$ (or equivalently $\nu$) we refer to as the principle parameter of the Morse potential, as it is this combination of the physical constants that determine the behaviour of the degeneracy in the spectrum. The eigenvectors of \eqref{eval} in their position representation are given by
\begin{equation}\label{mors}
\begin{split}
&\psi_{n,m}\left(x,y\right)=\bra{x,y}\ket{n,m}\\
&=\mathcal{N}_{n,m}e^{-\frac{\tilde{x}}{2}-\frac{\tilde{y}}{2}}\tilde{x}^{p-n}\tilde{y}^{p-m}L_n^{2(p-n)}\left( \tilde{x}\right)L_m^{2(p-m)}\left( \tilde{y}\right),
\end{split}
\end{equation}
where the tilde variables are related to the canonical position variables by
\begin{equation}
\tilde{x}=\nu e^{-\beta x}, \quad \tilde{y}=\nu e^{-\beta y}
\end{equation}
and the normalisation factor $\mathcal{N}_{n,m}$ is given explicitly by
\begin{equation}
\mathcal{N}_{n,m}=\beta \sqrt{\frac{(\nu-2n-1)(\nu-2m-1)\Gamma(n+1)\Gamma(m+1)}{\Gamma(\nu-n)\Gamma(\nu-m)}}.
\end{equation}
The 1D Morse potential admits only a finite number, $\floor{p}+1$, of bound states where $\floor{r}$ is the integer part of $r$. As such the two-dimensional Morse system admits $\left(\floor{p}+1\right)\times \left(\floor{p}+1\right)$ bound states. The quantum numbers $n,m$ take on the finite number of values
\begin{equation}
n,m\in \left\{0,1,\ldots,\floor{p} \right\}.
\end{equation}

\section{Analysis of the degeneracies of the energy spectrum}\label{sec3}
In this section we will focus on equation \eqref{energy} and the role of the principle parameter $p$ in determining the nature of the degeneracy in the spectrum. Assuming the principle parameter $p$ to be a real number, the structure of the degeneracy depends on the rationality of $p$. It is not possible to say whether $p$ is rational or not, it is built out of experimentally determined numbers $m, V_0, \beta, \hbar$ which themselves may be rational (or the number determined by experiment is a rational approximation), but the square root of the ratio \eqref{mainparam} may not be rational. We will demonstrate that accidental degeneracies are an inevitability when $p$ is taken to be rational, but they can be eliminated by choosing $p$ to be irrational.

Consider then the scaled bound state energy spectrum with parameter $p$ defined in \eqref{energy}
\begin{equation}\label{newe}
\begin{split}
\varepsilon_{n,m}=&-\left[\left( p-n\right)^2 +\left( p-m\right)^2\right],\\
&n,m\in \{0,1,\ldots,\floor{p}\}.
\end{split}
\end{equation}
Writing the parameter $p$ as the sum of its closest integer and a remainder
\begin{equation}
p=k+\epsilon,\quad k=\floor{p},\quad \epsilon \in [0,1),
\end{equation}
we may rewrite \eqref{newe} as
\begin{equation}\label{mainenergy}
\begin{split}
&\varepsilon_{n,m}(k,\epsilon)=\\
&-\left[\left( k-n\right)^2 +\left( k-m\right)^2+2\epsilon(2k-n-m)+2\epsilon^2\right].
\end{split}
\end{equation}
The three distinct cases in \eqref{mainenergy} we can discuss here are $p$ integer, $p$ rational and $p$ irrational. In the notation we have introduced these are:
\begin{align}
&\textrm{Case \RomanNumeralCaps{1}}& &\varepsilon_{n,m}\left(k,0\right) & &\textrm{$p$ integer}\\
&\textrm{Case \RomanNumeralCaps{2}}& &\varepsilon_{n,m}\left(k,\frac{r}{q}\right) & &\textrm{$p$ rational}\\
&\textrm{Case \RomanNumeralCaps{3}}& &\varepsilon_{n,m}(k,\epsilon) & &\textrm{$p$ irrational},
\end{align}
for $\frac{r}{q} \in [0,1) \subset\mathds{Q}$ and $\epsilon \in (0,1) \subset \mathds{R}\setminus\mathds{Q}$.
\subsection{Case I}
When the remainder term $\epsilon=0$ we find
\begin{equation}
\varepsilon_{n,m}(k,0)=-\left[\left( k-n\right)^2 +\left( k-m\right)^2\right].
\end{equation}
Solutions to this problem are well understood through Gaussian prime decomposition \cite{Shaw_1974}. As an example, if we take $k=9$
\begin{equation}
 \varepsilon_{n,m}(9,0)=-\left[\left( 9-n\right)^2 +\left( 9-m\right)^2\right],
\end{equation}
we find that
\begin{equation}\label{ex111}
 \varepsilon_{2,8}(9,0)= \varepsilon_{8,2}(9,0)= \varepsilon_{4,4}(9,0).
\end{equation}

In \eqref{ex111} the first two solutions are related by permuting the indices $n,m$, but neither are related to the third solution by any known symmetry. This problem is similar to that of the particle in a square box, though, because the quantum numbers $n,m$ take on only finitely many values, some solutions may be discarded. For instance $1^2+8^2=8^2+1^2=4^2+7^2=7^2+4^2$, but if $k=7$ then the solutions $(1,8), (8,1)$ lie outside of the bound state parameter range and as such should not be included as degenerate contributions.

To give a sense of scale to the problem of accidentally degenerate states, if we take the large example of $k=28$ then we have $841$ different bound states. After removing the doubly degenerate states (states symmetric under interchange of indices) we have $435$ states. Upon analysing the remaining states we find $360$ distinct values for the energy. This implies that there are $75$ accidentally degenerate (not counting the interchanging of their indices) states. The degrees of degeneracy also vary. Clearly this problem proliferates for larger $p$.
\subsection{Case II}
The next distinct case is for rational $p$, we may take the remainder term to be a rational number on the interval $[0,1)$
\begin{equation}
p\in \mathds{Q}, \quad \epsilon=\frac{r}{q}\in[0,1).
\end{equation}

The spectral problem then takes the form
\begin{equation}
\begin{split}
&\varepsilon_{n,m}\left(k,\frac{r}{q}\right)=\\
&-\left[\left( k-n\right)^2 +\left( k-m\right)^2+2\frac{r}{q}(2k-n-m) +2\frac{r^2}{q^2}\right].
\end{split}
\end{equation}
The rational $p$ case includes as a limiting case ($\epsilon=0$) the integer $p$ degeneracy problem but also a more general class of accidental degeneracies. Consider the following example of a rational $p$ degeneracy,
\begin{equation}
\begin{split}
 &\varepsilon_{n,m}\left(7,\frac{1}{2}\right)=\\
 &-\left[\left( 7-n\right)^2 +\left( 7-m\right)^2+(14-n-m) +\frac{1}{2}\right].
\end{split}
\end{equation}
we find
\begin{equation}
  \varepsilon_{2,6}\left(7,\frac{1}{2}\right)= \varepsilon_{6,2}\left(7,\frac{1}{2}\right)= \varepsilon_{3,4}\left(7,\frac{1}{2}\right)= \varepsilon_{4,3}\left(7,\frac{1}{2}\right).
\end{equation}
Clearly the first two solutions are related by a permutation of indices, as are the last two, but these two sets of solutions do not have a known symmetry connecting them. This generalises the previous problem where we found solutions represented as sum of two squares, this problem is the sum of two squares plus a fraction of the sum of the respective linear terms.

There is a subtle point to be made here, while accidental degeneracies can occur for rational $p$, it does not mean they will with certainty. If we want no accidental degeneracies to occur to simplify calculations, in order to implement the following results in a computer algebra system, it is necessary to find a rational value of $p$ close enough to our initial $p$ which does not produce accidental degeneracies. This usually means keeping more terms in the decimal expansion. For small enough values of $p$ it is straight forward to determine how many accidental degeneracies occur with a computer and thus it is easy to verify whether a certain choice of rational $p$ works well, though it is hard to make more general comments for arbitrarily large values of rational $p$.
\subsection{Case III}
Lastly, the case in which we will focus our attention from here on out is when $p$ is irrational. For $\epsilon$ some irrational number on $[0,1)$ the spectrum \eqref{mainenergy} is
\begin{equation}\label{acci}
\begin{split}
&\varepsilon_{n,m}(k, \epsilon)=\\
&-\left[\left( k-n\right)^2 +\left( k-m\right)^2+2\epsilon(2k-n-m)+2\epsilon^2\right].
\end{split}
\end{equation}
This equation has no accidentally degenerate solutions, only degenerate solutions obtained from the permutation of the indices $n,m$.

Indeed, for accidentally degenerate solutions to \eqref{acci}, $(n,m)$ and $(n',m')$, we must satisfy the rational and irrational parts of the equation separately because an irrational multiple of a rational number cannot coincide with a rational number, that is to say
\begin{equation}\label{e1}
\left( k-n\right)^2 +\left( k-m\right)^2=\left( k-n'\right)^2 +\left( k-m'\right)^2,
\end{equation}
and
\begin{equation}\label{e2}
\left( k-n\right) +\left( k-m\right)=\left( k-n'\right)+\left( k-m'\right),
\end{equation} 
must be satisfied. These equations may be thought of in terms of two triangles with the same length hypotenuse and perimeters. In order for \eqref{e1} and \eqref{e2} to be simultaneously satisfied we square \eqref{e2} and substitute in \eqref{e1}, in doing so imply the area equation for the triangles
\begin{equation}\label{area}
\left( k-n\right)\left( k-m\right)=\left( k-n'\right)\left( k-m'\right).
\end{equation}
Using the fact that $\left( k-n\right)=\left( k-n'\right)+\left( k-m'\right)-\left( k-m\right)$, substitution into \eqref{area} gives a quadratic equation in $(k-m)$
\begin{equation}
\begin{split}
\left( k-m\right)^2&-\left[\left( k-n'\right)+\left( k-m'\right)\right]\left( k-m\right)\\
&+\left( k-n'\right)\left( k-m'\right)=0
\end{split}
\end{equation}
which admits the solutions $m=n'$ or $m= m'$, after which $n$ is uniquely determined by \eqref{e2}. Thus we have at most doubly degenerate eigenvalues in the spectrum and these are precisely the degeneracies found by permuting the indices $n,m$. This makes the degeneracy problem more tractable. Irrationality of the principle parameter $p$ is the key to breaking the degeneracy symmetry.

\section{Ordering of the 2D non-degenerate spectrum}\label{sec4}
After discussing the degeneracies we are in a position to be able to organise the spectrum of the full 2D system in terms of a single index. If we consider the principle parameter $p$ to be irrational, we found that states are at most doubly degenerate. To this end whenever we encounter a doubly-degenerate eigenvalue we define the cumulative wavefunction with a pair of complex coefficients, $\gamma, \delta$, such that 

\begin{equation}\label{coeff}
\ket{\mu_k}_p=\gamma \ket{n,m} + \delta\ket{m,n}, \quad n>m, \quad \gamma,\delta \in \mathds{C}.
\end{equation}
Note that $n>m$ ensures that we uniformly introduce the coefficients throughout the spectrum. The complex coefficients \eqref{coeff} are subject to
\begin{equation}
\abs{\gamma}^2+\abs{\delta}^2=1,
\end{equation}
to preserve normalisation. Otherwise, for non degenerate states (states of the form $\ket{n,n}$) we simply take the definition
\begin{equation}
\ket{\mu_j}_p=\ket{n,n}.
\end{equation}
Considering the spectrum \eqref{acci}, where for convenience we remove the constant term, $\epsilon^2$, to define a shifted energy
\begin{equation}
\tilde{\varepsilon}_{n,m}(k, \epsilon)=-\left[\left( k-n\right)^2 +\left( k-m\right)^2+2\epsilon(2k-n-m)\right],
\end{equation}
which has maximum, $\max \tilde{\varepsilon}_{n,m}(k,\epsilon)=0$. We table values and order the energies as follows
\begin{equation}
\tilde{\varepsilon}_{0,0}(k, \epsilon)<\tilde{\varepsilon}_{1,0}(k, \epsilon)=\tilde{\varepsilon}_{0,1}(k, \epsilon)<\ldots<\tilde{\varepsilon}_{k,k}(k, \epsilon)=0
\end{equation}
once we remove duplicated energies, this inequality is in one to one correspondence with the single indexed spectrum
\begin{equation}
\varepsilon_{\mu_0}(k, \epsilon)<\varepsilon_{\mu_1}(k, \epsilon)<\ldots<\varepsilon_{\mu_\xi}(k, \epsilon).
\end{equation}
The number $\xi$ can be computed for any value of $k$, it is just the total number of unique elements in the $(k+1)\times (k+1)$ symmetric matrix with matrix elements $ \tilde{\varepsilon}_{n,m}(k,\epsilon)$ minus one (because we count the first state with index zero), it is given as
\begin{equation}\label{xidef}
\xi=\frac{(k+1)(k+2)}{2}-1.
\end{equation}
The final consideration we have is the dependence of the ordering on the parameter $\epsilon$. Take the example $k=3$ and the energies
\begin{equation}
\tilde{\varepsilon}_{3,0}(3, \epsilon)=9+6\epsilon, \quad \tilde{\varepsilon}_{1,1}(3,\epsilon)=8+8\epsilon,
\end{equation}
for $\epsilon<0.5$ we find $\tilde{\varepsilon}_{3,0}(3, \epsilon)>\tilde{\varepsilon}_{1,1}(3, \epsilon)$, while for $\epsilon >0.5$ we find $\tilde{\varepsilon}_{3,0}(3, \epsilon)<\tilde{\varepsilon}_{1,1}(3, \epsilon)$. This does prevent us for obtaining more general solutions for any irrational $\epsilon$. However, once $\epsilon$ is fixed, the ordering is uniquely determined.

\subsection{Non-degenerate states for $p=3 \pi$}
To illustrate the points made so far, we take the example of $p=3 \pi \approx 9.42478$, the bound state space is spanned by $100$ different eigenfunctions and there are $55$ distinct energy eigenvalues. This is our departing point, we will construct the set of states \eqref{coeff} which are associated to the distinct eigenvalues. In cases where doubly degenerate states appear we will introduce the parameters $\gamma,\delta$ that will control the mixing between the $x$ and $y$ modes to give one averaged contribution to the degenerate energy level.

The set
\begin{equation}\label{setofs}
\mathcal{S}=\left\{ \ket{\mu_0}_{3\pi}, \ket{\mu_1}_{3\pi}, \ldots, \ket{\mu_{54}}_{3\pi}\right\}, \quad \abs{\mathcal{S}}=55,
\end{equation}
provides all bound states of the problem under consideration. The states $ \ket{\mu_i}$ themselves are given in terms of the original eigenfunctions by
\begin{align}\label{nondeg}
\ket{\mu_i}_{3\pi}&=
\begin{cases}
\gamma\ket{n,m}+\delta\ket{m,n}, &\textrm{for }  n>m\\
\ket{n,n},  &\textrm{otherwise}.
\end{cases}
\end{align}
Explicitly, the first few and last states are
\begin{align}\label{states}
\ket{\mu_i}_{3\pi}&=
\begin{cases}
\ket{0,0}, & i=0\\
\gamma\ket{1,0}+\delta\ket{0,1}, & i=1\\
\gamma\ket{2,0}+\delta\ket{0,2}, & i=2\\
\ket{1,1}, & i=3\\
\vdots \\
\ket{9,9}, & i=54.
\end{cases}
\end{align}
The states $\ket{\mu_i}$ have energies which correspond to arranging the following energy function in increasing order
\begin{equation}\label{specex}
\begin{split}
&\varepsilon_{\mu_i}(9, 0.42478)=\\
&-\left[(9-n)^2 +(9-m)^2+2(0.42478)\left(18-n-m\right)\right],
\end{split}
\end{equation}
\begin{figure}
\centering
\includegraphics[scale=0.9]{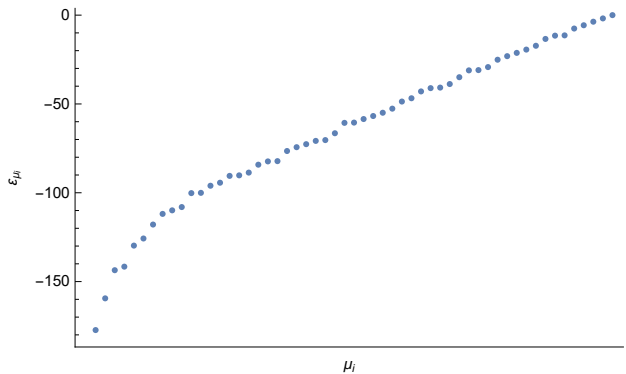}
\caption{Energies $\varepsilon_{\mu_0},\ldots,\varepsilon_{\mu_{54}}$ arranged in increasing order.}
\label{egraph}
\end{figure}
\begin{figure*}[!htb]
    \captionsetup[subfigure]{singlelinecheck=off}
\begin{tabularx}{\linewidth}{*{3}{X}}
\begin{subfigure}[b]{\linewidth}
\label{subfig:a}
\includegraphics[width=0.8\linewidth]{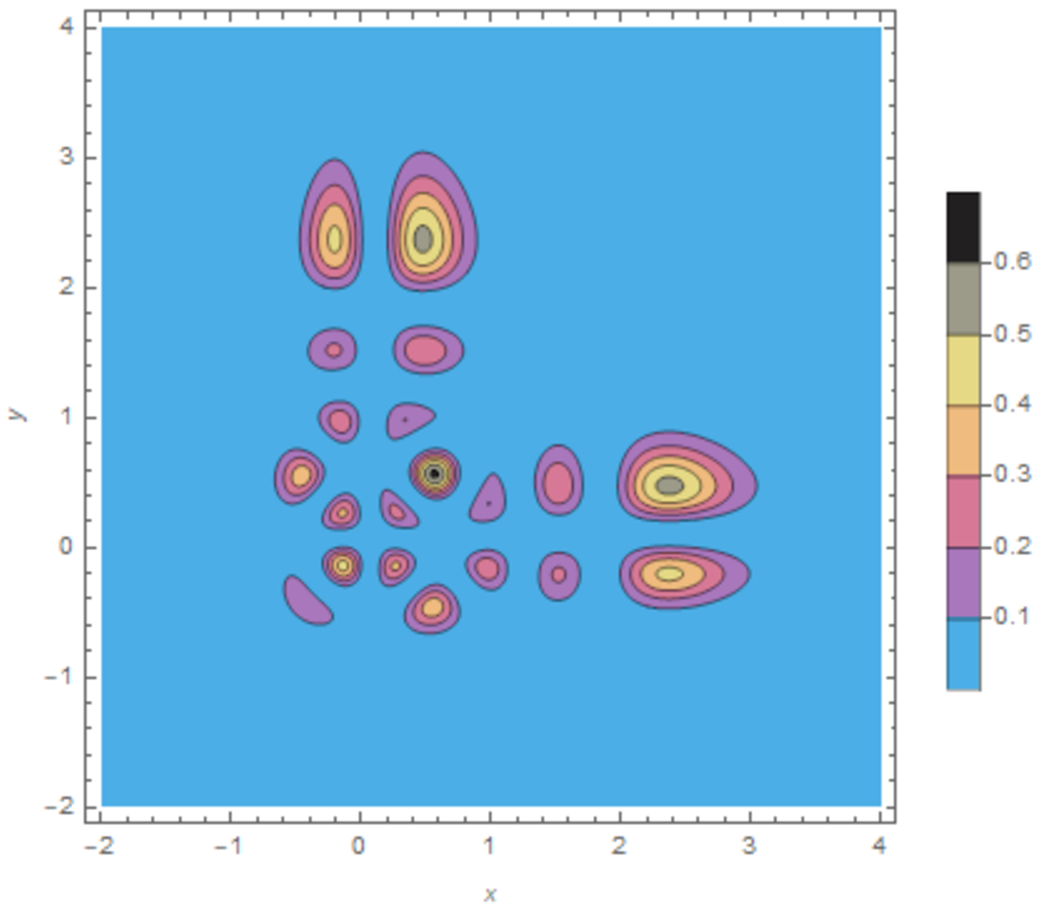}
\end{subfigure}
&
\begin{subfigure}[b]{\linewidth}
\label{subfig:b}
\includegraphics[width=0.8\linewidth]{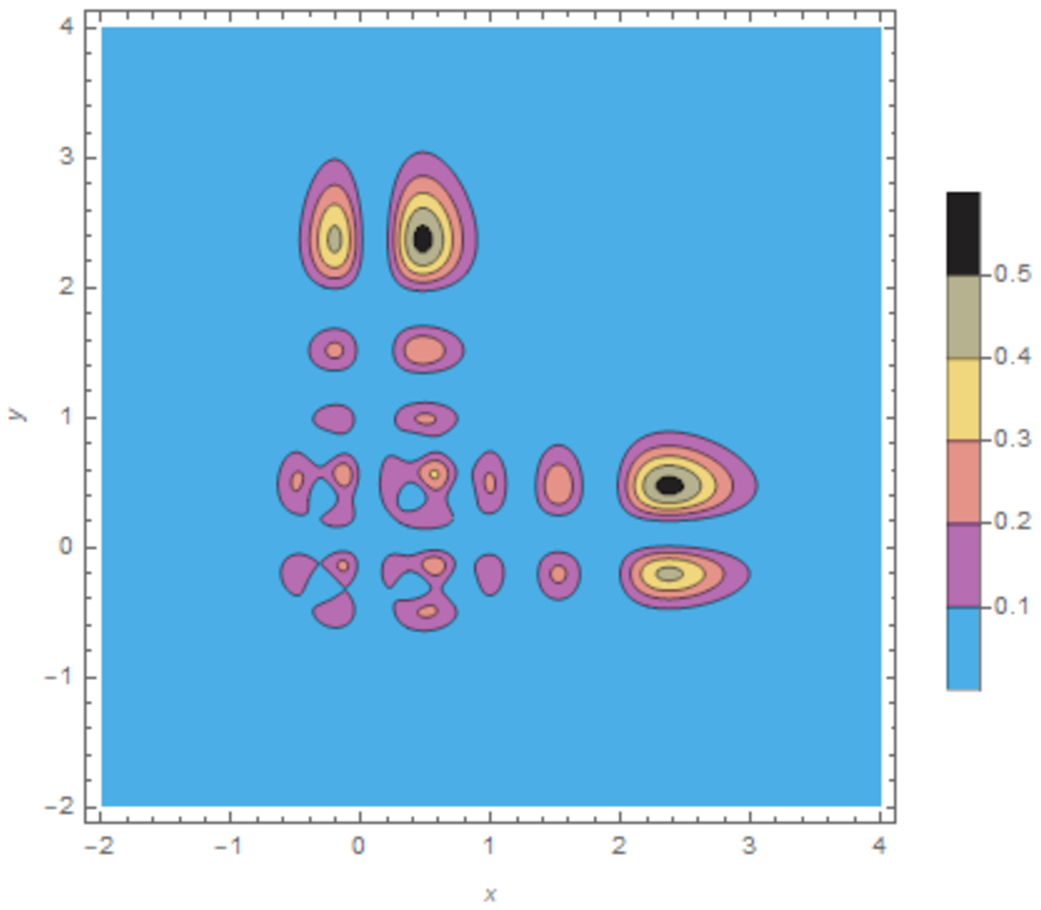}
\end{subfigure}
\end{tabularx}
    \caption{(Colour online) $ \abs{\bra{x,y}\ket{\mu_{18}}_{3\pi}}^2$, $p=3\pi$ with $\gamma=\frac{1}{\sqrt{2}},\delta=\frac{1}{\sqrt{2}}$ (left) and $\gamma=\frac{1}{\sqrt{2}}e^{i\frac{\pi}{2}},\delta=\frac{1}{\sqrt{2}}$ (right).}
    \label{fig:1}
\end{figure*}
\begin{figure*}[!htb]
    \captionsetup[subfigure]{singlelinecheck=off}
\begin{tabularx}{\linewidth}{*{3}{X}}
\begin{subfigure}[b]{\linewidth}
\label{subfig:a}
\includegraphics[width=0.8\linewidth]{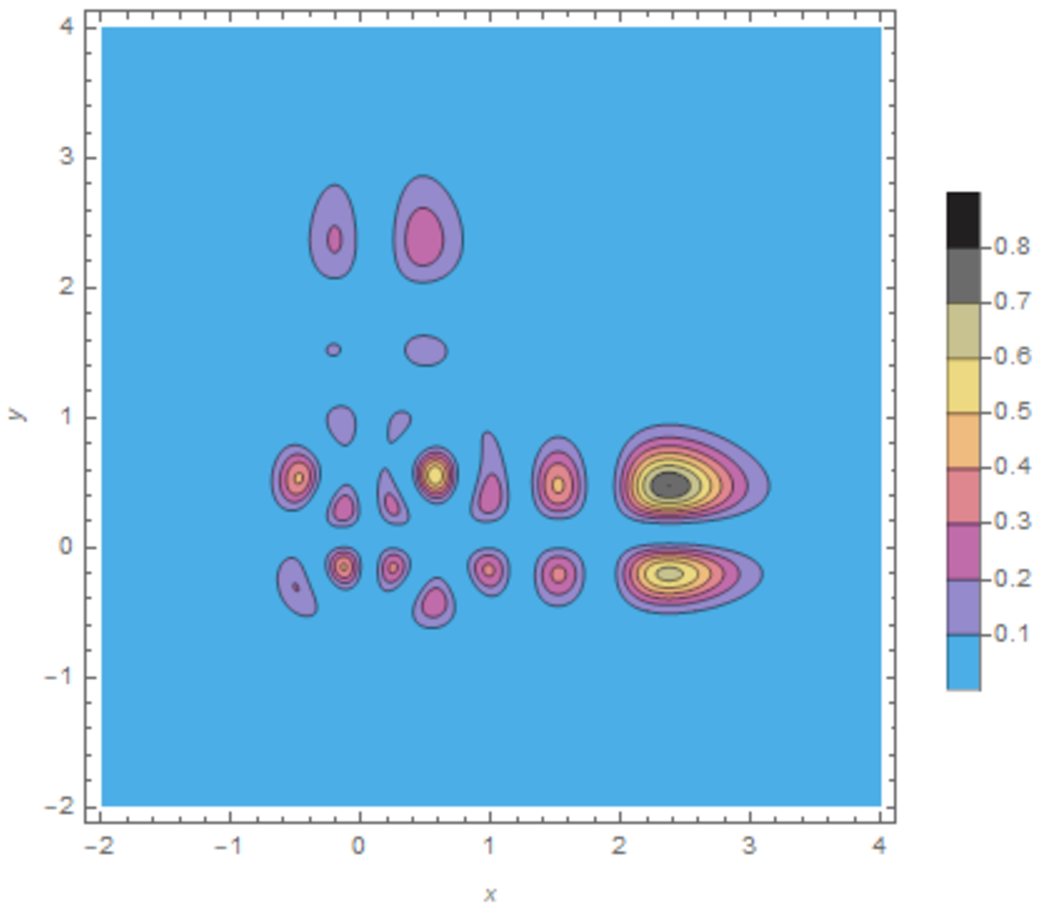}
\end{subfigure}
&
\begin{subfigure}[b]{\linewidth}
\label{subfig:b}
\includegraphics[width=0.8\linewidth]{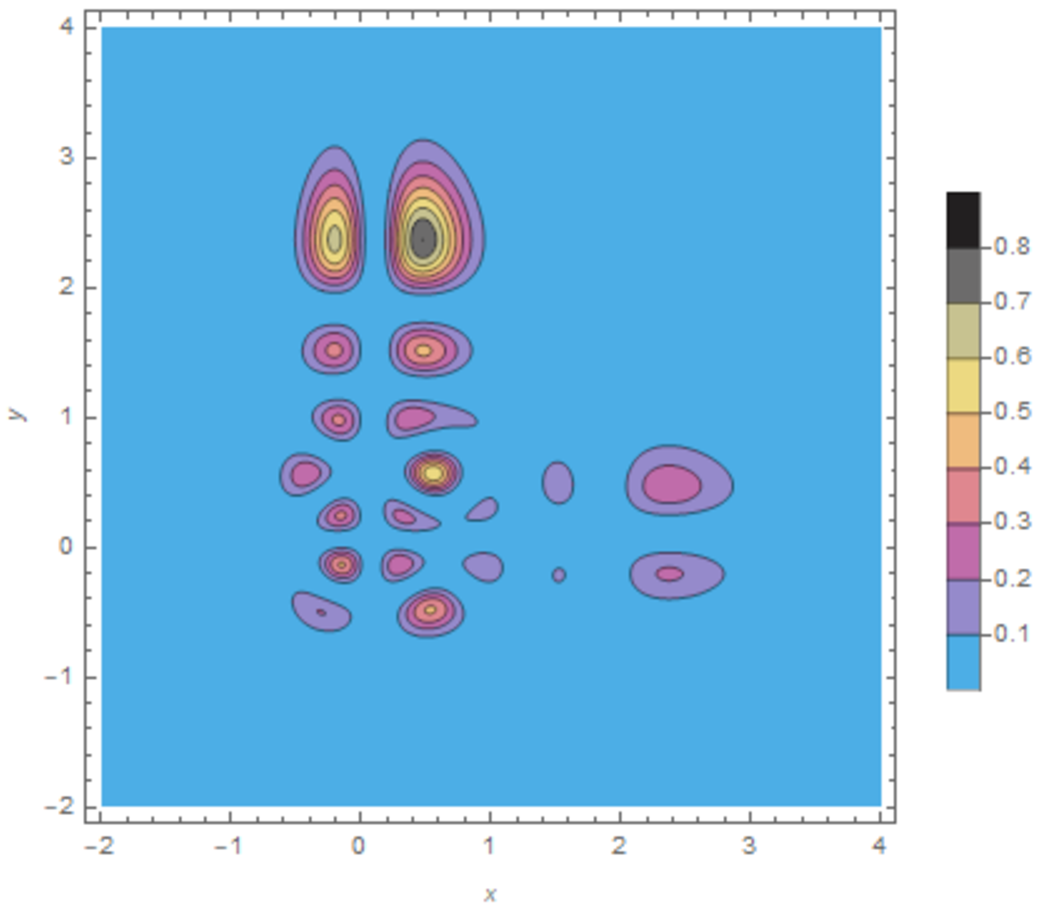}
\end{subfigure}
\end{tabularx}
    \caption{(Colour online) $ \abs{\bra{x,y}\ket{\mu_{18}}_{3\pi}}^2$, $p=3\pi$ with $\gamma=\frac{\sqrt{3}}{2},\delta=\frac{1}{2}$ (left) and $\gamma=\frac{1}{2},\delta=\frac{\sqrt{3}}{2}$ (right).}
    \label{fig:2}
\end{figure*}
The graph in figure \ref{egraph} shows the possible values that the energy function \eqref{specex} can take. These values are arranged in increasing order of magnitude. The states \eqref{nondeg} or \eqref{states} are in one to one correspondence with the spectrum \eqref{specex} and together they define a non-degenerate basis for the bound state sector of the 2D Morse potential at $p=3 \pi$. In figure \ref{fig:1} we see the effect of adding complex phase to a doubly degenerate state, $\ket{\mu_{18}}$, the phase alters the positioning of some of the `islands' of non-zero probability near to the origin due to the modes being in or out of phase, but it preserves the overall structure of the probability density. If on the other hand we change the magnitudes of $\gamma, \delta$ so that they are not equal we find that this corresponds to a larger change in the probability density function as seen in figure \ref{fig:2}. For $\gamma>\delta$ the higher energy $x$ mode is mixed with greater probability, and as a result most of the probability density occupies the islands which extend along the $x$ axis. Similarly, for $\delta>\gamma$, the most of the probability densities occupies the islands which extend along the $y$ axis.

\section{Ladder operators and coherent states}\label{sec5}
Coherent states for the harmonic oscillator were first studied by Schr\"odinger as minimal uncertainty wavepackets \cite{Schrodinger1926}. They may be defined through several equivalent means: as eigenstates of the annihilation operator, the action of the unitary displacement operator on the vacuum, or by their Fock space expansion. The three equivalent harmonic oscillator coherent state definitions read
\begin{align}\label{fockgen}
a^- \ket{\alpha}=\alpha \ket{\alpha}, \qquad & \textrm{Barut-Girardello}\\
\exp\left(\alpha a^+ - \bar{\alpha}a^- \right)\ket{0}=\ket{\alpha}, \qquad & \textrm{Displacement operator}\\\label{dada}
\ket{\alpha}=e^{-\frac{\abs{\alpha}^2}{2}}\sum_{n=0}^\infty \frac{\alpha^n}{\sqrt{n!}}\ket{n}, \qquad & \textrm{Fock expansion}.
\end{align}
Typically, for systems other than the harmonic oscillator, the three definitions do not coincide. Another consequences of these definitions is that the coherent states are minimal uncertainty with respect to the Heisenberg uncertainty relation with equal uncertainty in the position and momentum quadratures. In units of $\hbar=1$ this is
\begin{equation}
\left(\Delta \hat{Q}\right)^2_{\ket{\alpha}}\left(\Delta \hat{P}\right)^2_{\ket{\alpha}}= \frac{1}{4}, \quad \Delta \hat{Q}=\Delta \hat{P}.
\end{equation}
For some more general classes of coherent states, such as the squeezed states, the condition that the uncertainty in the position and momentum quadratures be equal is relaxed but their product still minimises the uncertainty relation.

When defining generalised coherent states for systems other than the harmonic oscillator, if we are not using any definitions regarding ladder operators or displacement operators, we use extensions of the form of \eqref{dada}. The extensions are formed by replacing the Fourier coefficients with some set of coefficients which satisfy completeness relations and produce good localisation of the coherent state wavefunction. As well, we replace the Fock basis vectors with the Fock states of the system under consideration. These definitions exist typically for 1D systems with non-degenerate spectra. In our case we used the preceding section to define a non-degenerate spectrum for the 2D Morse oscillator and as such allow ourselves to use these definitions.

Using the set of non-degenerate states
\begin{equation}
\mathcal{S}=\left\{ \ket{\mu_0}_{p}, \ket{\mu_1}_p, \ldots, \ket{\mu_{\xi}}_p\right\}, \quad \abs{\mathcal{S}}=\xi+1,
\end{equation}
and defining the set of ladder operators, $\mathcal{B}^+, \mathcal{B}^-$, such that
\begin{equation}
\begin{split}
&\mathcal{B}^+\ket{\mu_i}_{p}=\sqrt{f(i+1)}\ket{\mu_{i+1}}_{p},\\
 &\mathcal{B}^-\ket{\mu_i}_{p}=\sqrt{f(i)}\ket{\mu_{i-1}}_{p},
\end{split}
\end{equation}
subject to boundary conditions on the finite set of states, i.e. $f(0)=f(\xi+1)=0$. We can define a generalised coherent state from a Barut-Girardello type coherent state as an approximate eigenstate of $\mathcal{B}^-$ with complex eigenvalue $\Psi$,
\begin{equation}\label{bgcs}
\mathcal{B}^-\ket{\Psi}_{p}\approx\Psi\ket{\Psi}_{p}.
\end{equation}
Expanding $\ket{\Psi}_{p}$ in the basis $\ket{\mu_i}_{p}$ we retrieve the well known generalised coherent states \cite{PhysRevA.64.013817}
\begin{equation}\label{gencs}
\begin{split}
\ket{\Psi}_p=&\frac{1}{\sqrt{\mathcal{N} (\Psi)}}\sum_{n=0}^{\xi}\frac{\Psi^n}{\sqrt{[f(n)]!}}\ket{\mu_n}_{p},\\
&[f(n)]!=\prod_{m=0}^n f(m),
\end{split}
\end{equation}
where the generalised factorial takes the usual definition $[f(0)]!=1$ and the normalisation function
\begin{equation}
\mathcal{N} (\Psi)=\sum_{n=0}^{\xi} \frac{\abs{\Psi}^{2n}}{\left[f(n)\right]!},
\end{equation}
ensures $\prescript{}{p}{\bra{\Psi}\ket{\Psi}_p}=1$. The reason this is an approximate eigenstate is because of the finite spectrum, the last term in the expansion \eqref{gencs}, $\frac{\Psi^\xi}{\sqrt{[f(\xi)]!}}\ket{\mu_\xi}$, does not appear when computing \eqref{bgcs}. Nevertheless this term is typically very small and therefore does not contribute much, but by including it we can recover the finite spectrum version of the generalised coherent states defined in \cite{Klauder_1996}.

There is some freedom in the choice of function $f(i)$, but a natural choice in analogy to the harmonic oscillator is to use the difference in energies with respect to the ground states, that is
\begin{align}
f(i)=\begin{cases}\varepsilon_{\mu_i}(k, \epsilon)-\varepsilon_{\mu_0}(k, \epsilon), &\textrm{for } i\in\{0,1,\ldots,\xi\} \\
0, &\textrm{otherwise}.
\end{cases}
\end{align}
This function by definition satisfies the boundary conditions on the set, it appropriately annihilates the highest and lowest weight states. We will use this definition of coherent states from here on out.

\subsection{Application to $p=3 \pi$}
We now apply the formalism to our working example of $p=3\pi$. Using the set of 55 states defined in \eqref{setofs} along with spectrum \eqref{specex} we write the generalised coherent state \eqref{gencs} as
\begin{equation}\label{gencos}
\ket{\Psi}_{3\pi}=\frac{1}{\sqrt{\mathcal{N} (\Psi)}}\sum_{n=0}^{54}\frac{\Psi^n}{\sqrt{[\varepsilon_{\mu_n}-\varepsilon_{\mu_0}]!}}\ket{\mu_n}
\end{equation}
\begin{figure*}[!htb]
    \captionsetup[subfigure]{singlelinecheck=off}
\begin{tabularx}{\linewidth}{*{3}{X}}
\begin{subfigure}[b]{\linewidth}
\label{subfig:a}
\includegraphics[width=0.8\linewidth]{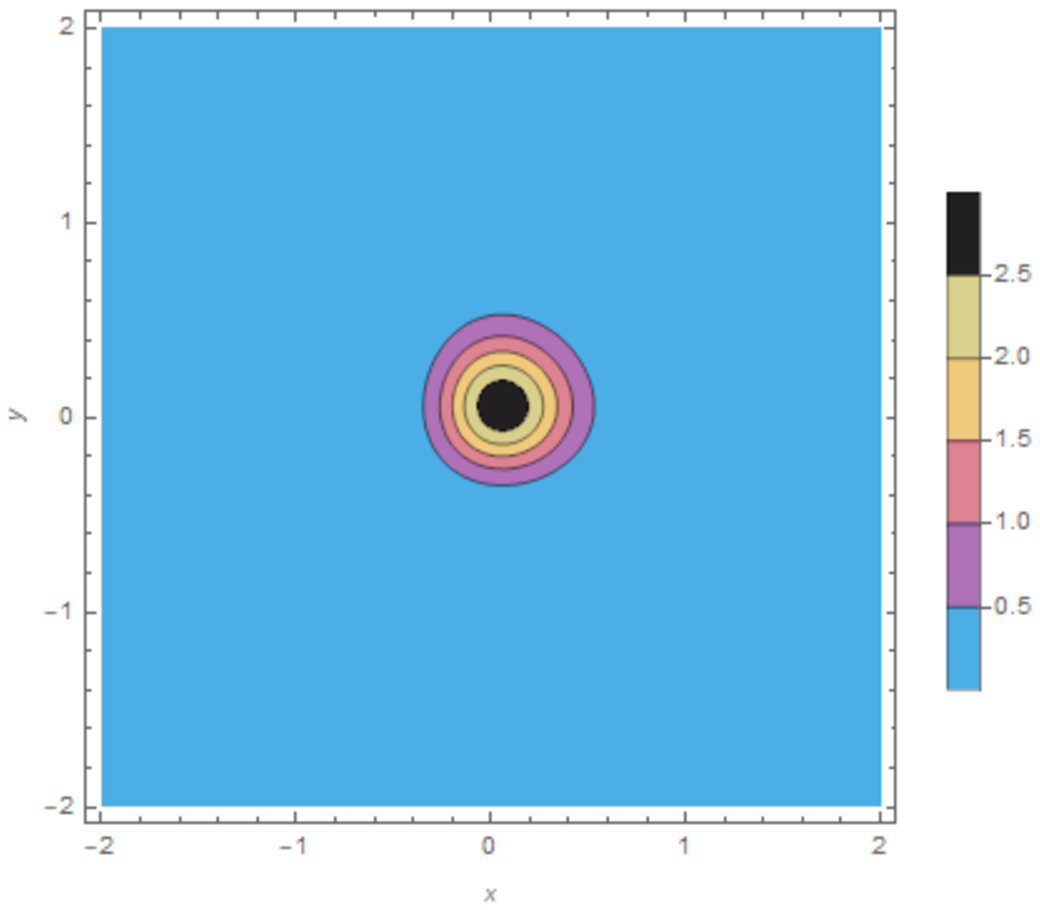}
\end{subfigure}
&
\begin{subfigure}[b]{\linewidth}
\label{subfig:b}
\includegraphics[width=0.8\linewidth]{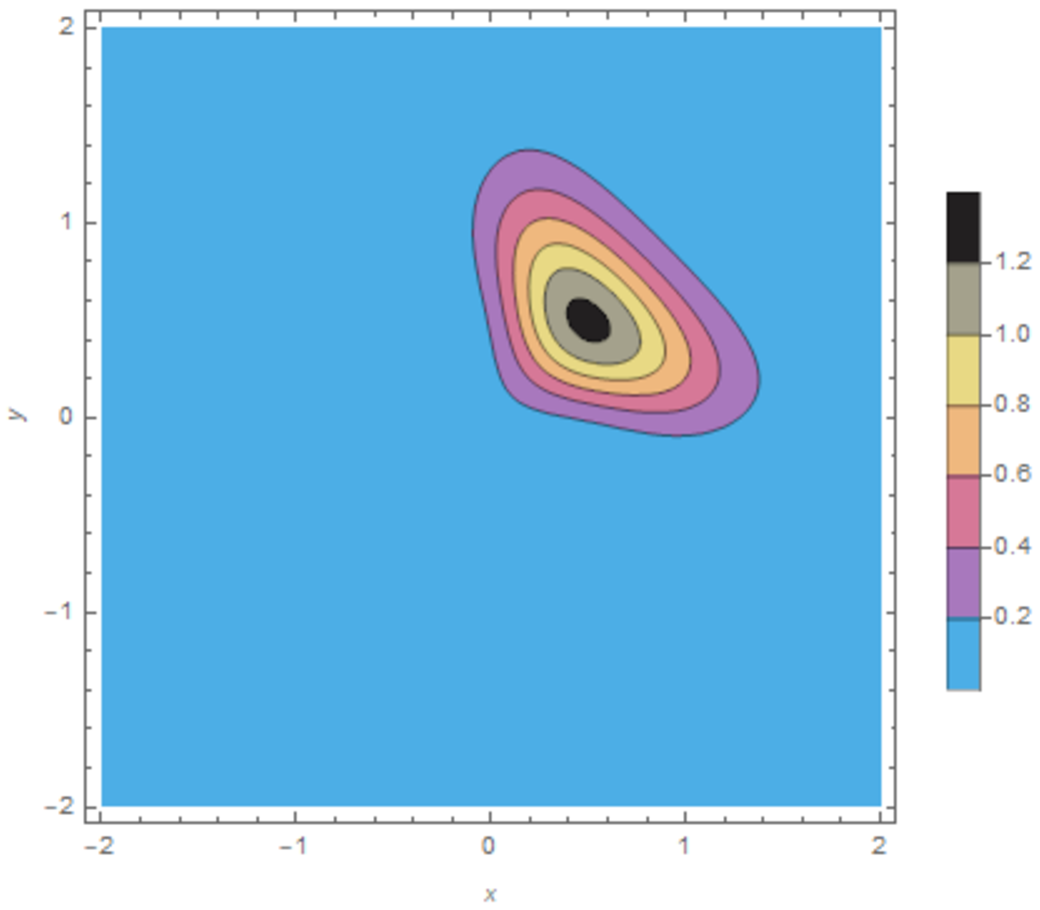}
\end{subfigure}
\end{tabularx}
    \caption{(Colour online) Spatial probability densities for the generalised coherent states, $\abs{\bra{x,y}\ket{\Psi}}^2$, at $\Psi=0.1$ (left) and $\Psi=5$ (right). Both with $\gamma=\frac{1}{\sqrt{2}},\delta=\frac{1}{\sqrt{2}}$.}
    \label{fig:example11}
\end{figure*}
\begin{figure*}[!htb]
    \captionsetup[subfigure]{singlelinecheck=off}
\begin{tabularx}{\linewidth}{*{3}{X}}
\begin{subfigure}[b]{\linewidth}
\label{subfig:a}
\includegraphics[width=0.8\linewidth]{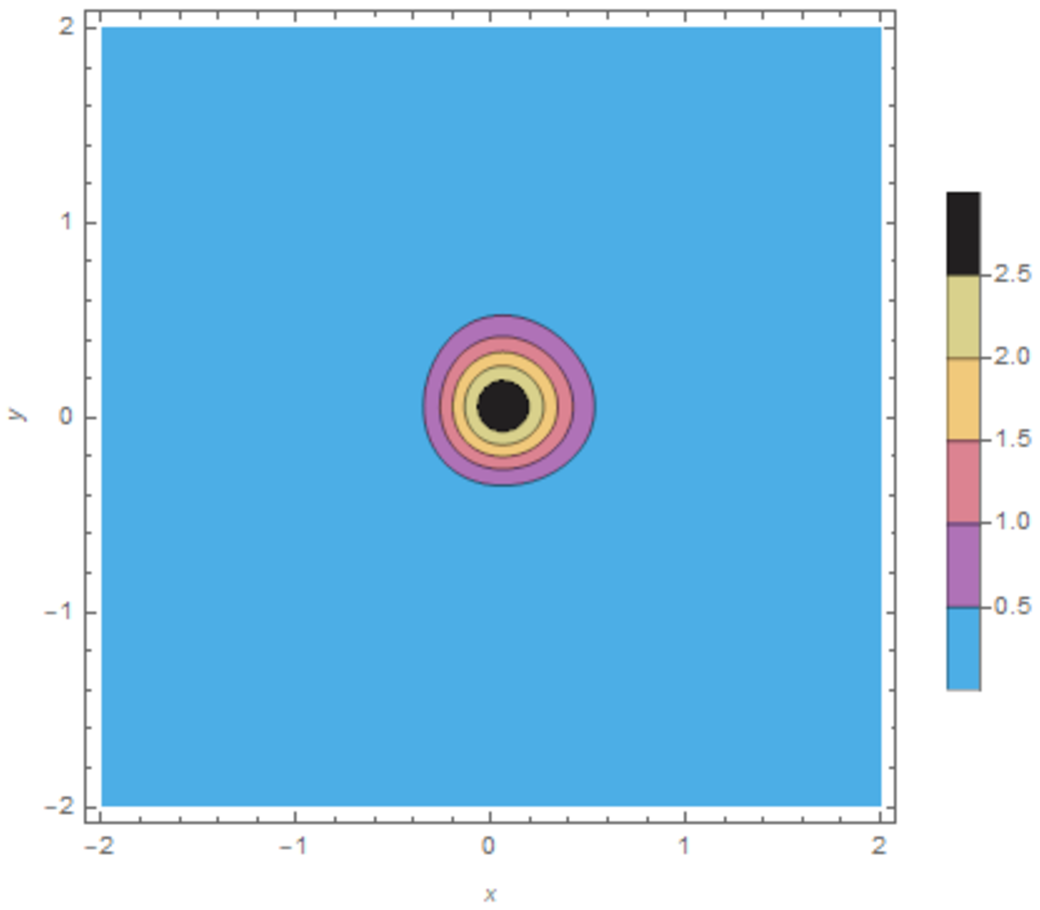}
\end{subfigure}
&
\begin{subfigure}[b]{\linewidth}
\label{subfig:b}
\includegraphics[width=0.8\linewidth]{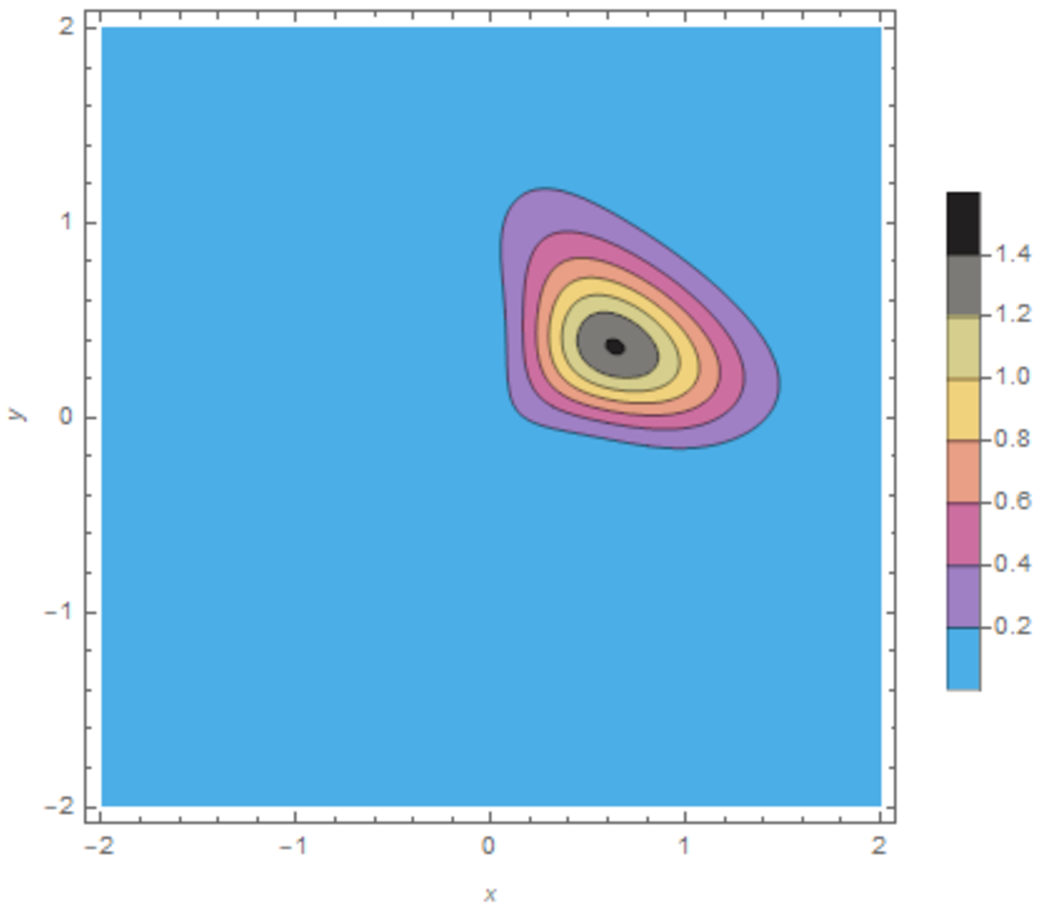}
\end{subfigure}
\end{tabularx}
    \caption{(Colour online) Spatial probability densities for the generalised coherent states, $\abs{\bra{x,y}\ket{\Psi}}^2$, at $\Psi=0.1$ (left) and $\Psi=5$ (right). Both with $\gamma=\frac{\sqrt{3}}{2},\delta=\frac{1}{2}$.}
    \label{fig:example1122}
\end{figure*}
When we consider $\Psi \in \mathds{R}^+$ we find that the wavefunction has better localisation for smaller $\Psi$. This is expected, because the generalised coherent states are formed by a power series in $\Psi$, when $\Psi$ is small, higher powers in the series contribute little to the wavefunction and the dominant contribution is the ground state.

The effect of the parameters $\gamma, \delta$ is minimal when the coherent state parameter, $\Psi$, is suitably chosen. If we set $\gamma \not = \delta$ we do induce some asymmetry about the line $y=x$ in the probability distribution as seen in figure \ref{fig:example1122}. This effect is more apparent for larger values of $\Psi$ and almost indistinguishable from the $\gamma=\delta$ case for small values of $\Psi$. The coherent states are most sensitive to changes in $\Psi$, but we have additional control over their behaviour by adjusting $\gamma, \delta$.

For the generalised coherent states \eqref{gencos} we find that the Heisenberg uncertainty relation is satisfied, moreover, it is closer to its minimum for smaller values of $\Psi$ as expected.
\begin{equation}
\left(\Delta \hat{Q}_s\right)^2_{\ket{\Psi}}\left(\Delta \hat{P}_s\right)^2_{\ket{\Psi}}\geq \frac{1}{4}, \quad s=x,y.
\end{equation}

\begin{figure*}[!htb]
\centering
\includegraphics[scale=1]{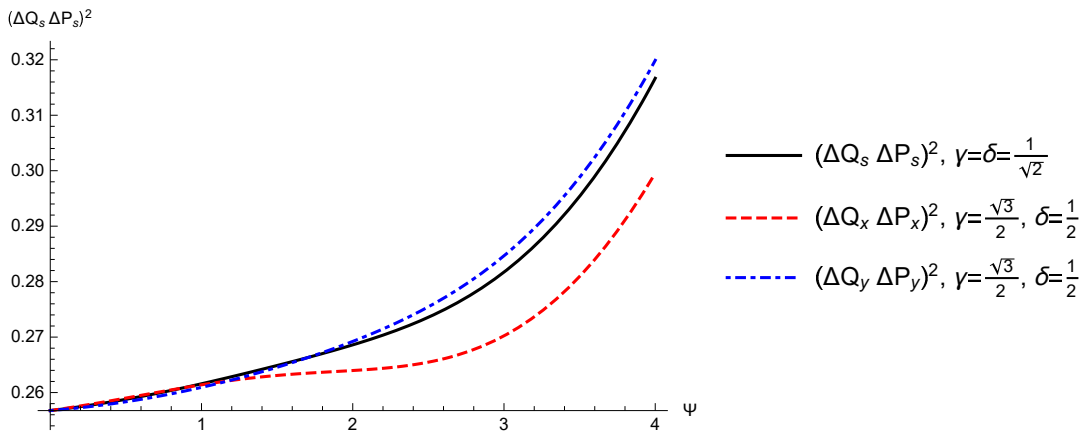}
\caption{(Colour online) Uncertainty relations in the $x$ and $y$ modes for the generalised coherent states in the symmetric, $\gamma=\delta=\frac{1}{\sqrt{2}}$, and asymmetric, $\gamma=\frac{\sqrt{3}}{2},\delta=\frac{1}{2}$,  regimes. Here $s=x,y$.}
\label{graph}
\end{figure*}

In figure \ref{graph} we plot the product uncertainty relations in the $x$ and $y$ modes with equal and unequal values of $\gamma, \delta$ for the generalised coherent states as a function of $\Psi$. We see that for the generalised coherent states the product of the uncertainties remains close to the minimum for values $\Psi<1$, and the effect of asymmetry between $\gamma, \delta$ is minimal. For larger $\Psi$ however, the wavefunctions begin to delocalise and we also observe a growth in the product of uncertainties. This is also reflected in the spatial distribution in figures \ref{fig:example11} and \ref{fig:example1122}. In the case where the parameters $\gamma, \delta$ are equal we do not introduce any asymmetry between the two modes and thus the uncertainty relations look identical for both the $x$ and $y$ modes.

When we include some asymmetry by setting the parameters $\gamma > \delta$, we find that the product of the uncertainties for the generalised coherent states in the $x$ mode are smaller than those of the $y$ mode when $\Psi$ is large enough. This effect is most noticeable for $\Psi>1.4$. Again the parameters $\gamma,\delta$ offer additional control over the behaviour of the generalised coherent states. We can, in effect, reduce the product uncertainties in one mode at the expense of increasing the product uncertainties in the other mode. The global behaviour is still determined by the coherence parameter $\Psi$.

\section{Conclusion and outlook}\label{sec6}
We found a scheme for constructing a singled indexed set of non-degenerate states for the 2D Morse potential. We assessed three distinct forms the spectrum can take corresponding to the rationality of the principle parameter $p$ and discussed their degeneracies. The critical observation is that the irrationality of the principle parameter $p$ implies that the degeneracy in the 2D Morse potential is at most two-fold. This follows from a straightforward analysis of the spectrum. In restricting to irrational choices of $p$ we make the problem of handling degeneracy in the system much more tractable, and correspondingly we only need to introduce two meaningful complex parameters, $\gamma, \delta$, subject to a normalisation constraint in order to define a degeneracy free spectrum for the 2D system. The solution we have used is algorithmic in approach and the techniques discussed here should be applicable to any two-dimensional system with quadratically degenerate spectra.

We saw that the introduction of the parameters $\gamma, \delta$ serve to tune the concentration of the probability densities of the non-degenerate states in configuration space by controlling the weight of the contributing $x$ and $y$ modes. Furthermore, we introduced Barut-Girardello type generalised coherent states from a set of ladder operators acting on the non-degenerate set of states and found that they are well localised in their spatial distribution and approximately minimise the Heisenberg uncertainty relation for small values of the coherence parameter $\Psi$. Additionally, we found the effect of $\gamma, \delta$ to be more significant in the coherent states for larger values of $\Psi$.
 
We relied on an algorithmic method to deal with the degeneracy problem, but algebraic and symmetry approaches may offer further insight into the structure of the degeneracies. Finally, it would also be interesting to study the sets of states we could generate for the supersymmetric partners of the 2D Morse system whose potential functions are non-separable, and their respective coherent states under this construction.

\begin{acknowledgments}
J. Moran acknowledges the support of the Département de physique at the Universit\'e de Montr\'eal. J. Moran would also like to thank V. Hussin and I. Marquette for their helping in preparing this manuscript. 
\end{acknowledgments}

\bibliography{refs}

\end{document}